\documentstyle[12pt]{article}

\skewchar\fivmi='177
\skewchar\sixmi='177
\skewchar\sevmi='177
\skewchar\egtmi='177
\skewchar\ninmi='177
\skewchar\tenmi='177
\skewchar\elvmi='177
\skewchar\twlmi='177
\skewchar\frtnmi='177
\skewchar\svtnmi='177
\skewchar\twtymi='177
\def\@magscale#1{ scaled \magstep #1}
\skewchar\fivsy='60
\skewchar\sixsy='60
\skewchar\sevsy='60
\skewchar\egtsy='60
\skewchar\ninsy='60
\skewchar\tensy='60
\skewchar\elvsy='60
\skewchar\twlsy='60
\skewchar\frtnsy='60
\skewchar\svtnsy='60
\skewchar\twtysy='60

\catcode`@=11
\def\un#1{\relax\ifmmode\@@underline#1\else
        $\@@underline{\hbox{#1}}$\relax\fi}
\catcode`@=12

\def\a{\alpha}
\def\b{\beta}

\def\d{\delta}
\def\e{\epsilon}

\def\l{\lambda}

\def\p{\pi}

\def\r{\rho}
\def\s{\sigma}

\def\z{\zeta}

\def\P{\Pi}

\def\U{\Upsilon}

\def\EQ{\begin{equation}}
\def\EN{\end{equation}}
\def\bea{\begin{eqnarray}}
\def\ena{\end{eqnarray}}

\def\beq{\begin{equation}}
\def\eeq{\end{equation}}
\def\bea{\begin{eqnarray}}
\def\eea{\end{eqnarray}}

\def\Mb{\kern 2pt\mathchoice
            {
             \vbox{\hrule width10pt height 0.4pt depth 0pt
                 \kern 1.2pt\hbox{\kern -2pt$\displaystyle M$}}}
            {
                 \vbox{\hrule width10pt height 0.4pt depth 0pt
                 \kern 1.2pt\hbox{\kern -2pt$\textstyle M$}}}
            {
\vbox{\hrule width6pt height 0.4pt depth 0pt
                 \kern 1.0pt\hbox{\kern -2pt$\scriptstyle M$}}}
            {
                 \vbox{\hrule width5pt height 0.4pt depth 0pt
                 \kern 0.8pt\hbox{\kern -2pt$\scriptscriptstyle M$}}}}

\def\Sb{\kern 2pt\mathchoice
            {
                 \vbox{\hrule width6pt height 0.4pt depth 0pt
                 \kern 1.2pt\hbox{\kern -2pt$\displaystyle S$}}}
            {
                 \vbox{\hrule width6pt height 0.4pt depth 0pt
                 \kern 1.2pt\hbox{\kern -2pt$\textstyle S$}}}
            {
                 \vbox{\hrule width3.5pt height 0.4pt depth 0pt
                 \kern 1.0pt\hbox{\kern -2pt$\scriptstyle S$}}}
            {
                 \vbox{\hrule width3pt height 0.4pt depth 0pt
                 \kern 0.8pt\hbox{\kern -2pt$\scriptscriptstyle S$}}}}

\def\Rb{\kern 2pt\mathchoice
            {
                 \vbox{\hrule width5.5pt height 0.4pt depth 0pt
                 \kern 1.2pt\hbox{\kern -2.5pt$\displaystyle R$}}}
            {
                 \vbox{\hrule width5.5pt height 0.4pt depth 0pt
                 \kern 1.2pt\hbox{\kern -2.5pt$\textstyle R$}}}
            {
                 \vbox{\hrule width3.5pt height 0.4pt depth 0pt
                 \kern 1.0pt\hbox{\kern -2.2pt$\scriptstyle R$}}}
            {
                 \vbox{\hrule width3pt height 0.4pt depth 0pt
                 \kern 0.8pt\hbox{\kern -2.2pt$\scriptscriptstyle R$}}}}

  \def\pp{{\mathchoice
          {
              \kern 1pt%
              \raise 1pt
              \vbox{\hrule width5pt height0.4pt depth0pt
                    \kern -2pt
                    \hbox{\kern 2.3pt
                          \vrule width0.4pt height6pt depth0pt
                          }
                    \kern -2pt
                    \hrule width5pt height0.4pt depth0pt}%
                    \kern 1pt
           }
            {
              \kern 1pt%
              \raise 1pt
              \vbox{\hrule width4.3pt height0.4pt depth0pt
                    \kern -1.8pt
                    \hbox{\kern 1.95pt
                          \vrule width0.4pt height5.4pt depth0pt
                          }
                    \kern -1.8pt
                    \hrule width4.3pt height0.4pt depth0pt}%
                    \kern 1pt
            }
            {
              \kern 0.5pt%
              \raise 1pt
              \vbox{\hrule width4.0pt height0.3pt depth0pt
                    \kern -1.9pt  
                    \hbox{\kern 1.85pt
                          \vrule width0.3pt height5.7pt depth0pt
                          }
                    \kern -1.9pt
                    \hrule width4.0pt height0.3pt depth0pt}%
                    \kern 0.5pt
            }
            {
              \kern 0.5pt%
              \raise 1pt
              \vbox{\hrule width3.6pt height0.3pt depth0pt
                    \kern -1.5pt
                    \hbox{\kern 1.65pt
                          \vrule width0.3pt height4.5pt depth0pt
                          }
                    \kern -1.5pt
                    \hrule width3.6pt height0.3pt depth0pt}%
                    \kern 0.5pt
            }
        }}

  \def\mm{{\mathchoice
                       {
                             \kern 1pt
               \raise 1pt    \vbox{\hrule width5pt height0.4pt depth0pt
                                  \kern 2pt
                                  \hrule width5pt height0.4pt depth0pt}
                             \kern 1pt}
                       {
                            \kern 1pt
               \raise 1pt \vbox{\hrule width4.3pt height0.4pt depth0pt
                                  \kern 1.8pt
                                  \hrule width4.3pt height0.4pt depth0pt}
                             \kern 1pt}
                       {
                            \kern 0.5pt
               \raise 1pt
                            \vbox{\hrule width4.0pt height0.3pt depth0pt
                                  \kern 1.9pt
                                  \hrule width4.0pt height0.3pt depth0pt}
                            \kern 1pt}
                       {
                           \kern 0.5pt
             \raise 1pt  \vbox{\hrule width3.6pt height0.3pt depth0pt
                                  \kern 1.5pt
                                  \hrule width3.6pt height0.3pt depth0pt}
                           \kern 0.5pt}
                       }}

\def\pd{{\kern0.5pt
                   + \kern-5.05pt \raise5.8pt\hbox{$\textstyle.$}\kern
0.5pt}}

\def\pmd{{\kern0.5pt
                  \pm \kern-5.05pt \raise6.3pt\hbox{$\textstyle.$}\kern1.5pt}}

\def\md{{\mathchoice
   {
      {{\kern 1pt - \kern-6.2pt \raise5pt\hbox{$\textstyle.$}\kern 1pt}}}
    {
      {{\kern 1pt - \kern-6.2pt \raise5pt\hbox{$\textstyle.$}\kern 1pt}}}
    {
      {\kern0.5pt - \kern-5.05pt \raise3.4pt\hbox{$\textstyle.$}\kern0.5pt}}
    {
      {\kern0.5pt - \kern-5.05pt \raise3.4pt\hbox{$\textstyle.$}\kern0.5pt}}}}

\def\pp{{\mathchoice
          {
              \kern 1pt%
              \raise 1pt
              \vbox{\hrule width5pt height0.4pt depth0pt
                    \kern -2pt
                    \hbox{\kern 2.3pt
                          \vrule width0.4pt height6pt depth0pt
                          }
                    \kern -2pt
                    \hrule width5pt height0.4pt depth0pt}%
                    \kern 1pt
           }
            {
              \kern 1pt%
              \raise 1pt
              \vbox{\hrule width4.3pt height0.4pt depth0pt
                    \kern -1.8pt
                    \hbox{\kern 1.95pt
                          \vrule width0.4pt height5.4pt depth0pt
                          }
                    \kern -1.8pt
                    \hrule width4.3pt height0.4pt depth0pt}%
                    \kern 1pt
            }
            {
              \kern 0.5pt%
              \raise 1pt
              \vbox{\hrule width4.0pt height0.3pt depth0pt
                    \kern -1.9pt  
                    \hbox{\kern 1.85pt
                          \vrule width0.3pt height5.7pt depth0pt
                          }
                    \kern -1.9pt
                    \hrule width4.0pt height0.3pt depth0pt}%
                    \kern 0.5pt
            }
            {
              \kern 0.5pt%
              \raise 1pt
              \vbox{\hrule width3.6pt height0.3pt depth0pt
                    \kern -1.5pt
                    \hbox{\kern 1.65pt
                          \vrule width0.3pt height4.5pt depth0pt
                          }
                    \kern -1.5pt
                    \hrule width3.6pt height0.3pt depth0pt}%
                    \kern 0.5pt
            }
        }}

  \def\mm{{\mathchoice
                       {
                             \kern 1pt
               \raise 1pt    \vbox{\hrule width5pt height0.4pt depth0pt
                                  \kern 2pt
                                  \hrule width5pt height0.4pt depth0pt}
                             \kern 1pt}
                       {
                            \kern 1pt
               \raise 1pt \vbox{\hrule width4.3pt height0.4pt depth0pt
                                  \kern 1.8pt
                                  \hrule width4.3pt height0.4pt depth0pt}
                             \kern 1pt}
                       {
                            \kern 0.5pt
               \raise 1pt
                            \vbox{\hrule width4.0pt height0.3pt depth0pt
                                  \kern 1.9pt
                                  \hrule width4.0pt height0.3pt depth0pt}
                            \kern 1pt}
                       {
                           \kern 0.5pt
             \raise 1pt  \vbox{\hrule width3.6pt height0.3pt depth0pt
                                  \kern 1.5pt
                                  \hrule width3.6pt height0.3pt depth0pt}
                           \kern 0.5pt}
                       }}

\def\pd{{\kern0.5pt
                   + \kern-5.05pt \raise5.8pt\hbox{$\textstyle.$}\kern
0.5pt}}

\def\pmd{{\kern0.5pt
                  \pm \kern-5.05pt \raise6.3pt\hbox{$\textstyle.$}\kern1.5pt}}

\def\md{{\mathchoice
   {
      {{\kern 1pt - \kern-6.2pt \raise5pt\hbox{$\textstyle.$}\kern 1pt}}}
    {
      {{\kern 1pt - \kern-6.2pt \raise5pt\hbox{$\textstyle.$}\kern 1pt}}}
    {
      {\kern0.5pt - \kern-5.05pt \raise3.4pt\hbox{$\textstyle.$}\kern0.5pt}}
    {
      {\kern0.5pt - \kern-5.05pt \raise3.4pt\hbox{$\textstyle.$}\kern0.5pt}}}}

\def\dslash{\not{\hbox{\kern-2pt $\partial$}}}
\def\Dslash{\not{\hbox{\kern-4pt $D$}}}
\def\pslash{\not{\hbox{\kern-2.3pt $p$}}}
 \newtoks\slashfraction
 \slashfraction={.13}
 \def\slash#1{\setbox0\hbox{$ #1 $}
 \setbox0\hbox to \the\slashfraction\wd0{\hss \box0}/\box0 }
 
\font\ro=cmsy10                          
\def\kcr{{\hbox{\ro \char'170}}}                
\def\ktl{{\hbox{\ro \char'170}}}        
\def\ktr{{\hbox{\ro \char'170}}}        
\def\kbl{{\hbox{\ro \char'170}}}        
\def\kbr{{\hbox{\ro \char'170}}}        

\def\plpl{\raise-2pt\hbox{$\raise3pt\hbox{$_+$}\hskip-6.67pt\raise0.0pt
\hbox{$^+$}\hskip 0.01pt$}}
\def\mimi{\raise-2pt\hbox{$\raise3pt\hbox{$_-$}\hskip-6.67pt\raise0.0pt
\hbox{$^-$}\hskip 0.01pt$}} 

\def\bo{{\raise.15ex\hbox{\large$\Box$}}}               
\def\pa{\partial}                                       
\def\TH{{\raise.2ex\hbox{$\displaystyle \bigodot$}\mskip-4.7mu \llap H \;}}
\def\face{{\raise.2ex\hbox{$\displaystyle \bigodot$}\mskip-2.2mu \llap {$\ddot
        \smile$}}}                                      

\def\sp#1{{}^{#1}}                              
\def\Tilde#1{\widetilde{#1}}                    
\def\Hat#1{\widehat{#1}}                        
\def\Bar#1{\overline{#1}}                       
\def\leftrightarrowfill{$\mathsurround=0pt \mathord\leftarrow \mkern-6mu
        \cleaders\hbox{$\mkern-2mu \mathord- \mkern-2mu$}\hfill
        \mkern-6mu \mathord\rightarrow$}
\def\dvec#1{\vbox{\ialign{##\crcr
        \leftrightarrowfill\crcr\noalign{\kern-1pt\nointerlineskip}
        $\hfil\displaystyle{#1}\hfil$\crcr}}}           

\def\frac#1#2{{\textstyle{#1\over\vphantom2\smash{\raise.20ex
        \hbox{$\scriptstyle{#2}$}}}}}                   
\def\sfrac#1#2{{\vphantom1\smash{\lower.5ex\hbox{\small$#1$}}\over
        \vphantom1\smash{\raise.4ex\hbox{\small$#2$}}}} 
\def\bfrac#1#2{{\vphantom1\smash{\lower.5ex\hbox{$#1$}}\over
        \vphantom1\smash{\raise.3ex\hbox{$#2$}}}}       
\def\afrac#1#2{{\vphantom1\smash{\lower.5ex\hbox{$#1$}}\over#2}}    

\def\bea{\begin{eqnarray}}
\def\eea{\end{eqnarray}}
\def\beq{\begin{equation}}
\def\eeq{\end{equation}}

\newskip\humongous \humongous=0pt plus 1000pt minus 1000pt
\def\caja{\mathsurround=0pt}
\def\eqalign#1{\,\vcenter{\openup2\jot \caja
        \ialign{\strut \hfil$\displaystyle{##}$&$
        \displaystyle{{}##}$\hfil\crcr#1\crcr}}\,}
\newif\ifdtup

\def\ref#1{$\sp{#1)}$}

\parskip=\medskipamount                 
\thicklines                         

\def\oldheadpic{                                
        \setlength{\unitlength}{.4mm}
        \thinlines
        \par
        \begin{picture}(349,16)
        \put(325,16){\line(1,0){4}}
        \put(330,16){\line(1,0){4}}
        \put(340,16){\line(1,0){4}}
        \put(335,0){\line(1,0){4}}
        \put(340,0){\line(1,0){4}}
        \put(345,0){\line(1,0){4}}
        \put(329,0){\line(0,1){16}}
        \put(330,0){\line(0,1){16}}
        \put(339,0){\line(0,1){16}}
        \put(340,0){\line(0,1){16}}
        \put(344,0){\line(0,1){16}}
        \put(345,0){\line(0,1){16}}
        \put(329,16){\oval(8,32)[bl]}
        \put(330,16){\oval(8,32)[br]}
        \put(339,0){\oval(8,32)[tl]}
        \put(345,0){\oval(8,32)[tr]}
        \end{picture}
        \par
        \thicklines
        \vskip.2in}
\def\oldtitle#1#2#3#4{\oldheadpic\begin{center}\vglue.5in{\large\bf #1}\\[.6in]
        {#2}\\[.1in] {\it Department of Physics and Astronomy}\\
        {\it University of Maryland, College Park, MD 20742}\\[.6in]
        Physics Publication \#{#3}\\ {#4}\\[1.5in] {\bf ABSTRACT}\\[.1in]
        \end{center} \begin{quotation}}                 
\def\oldTitle#1#2#3#4#5#6#7{\oldheadpic\begin{center} \vglue .4in
        {\large\bf #1}\\[.4in]
        {#2}\\[.1in] {\it Department of Physics and Astronomy}\\
        {\it University of Maryland, College Park, MD 20742}\\[.1in]
        {#3}\\[.1in] {\it {#4}}\\ {\it {#5}}\\[.4in]
        Physics Publication \#{#6}\\ {#7}\\[.5in] {\bf ABSTRACT}\\[.1in]
        \end{center} \begin{quotation}}                 
\def\border{                                            
        \setlength{\unitlength}{1mm}
        \newcount\xco
        \newcount\yco
        \xco=-21
        \yco=12
        \begin{picture}(140,0)
        \put(\xco,\yco){$\ktl$}
        \advance\yco by-1
        {\loop
        \put(\xco,\yco){$\kcr$}
        \advance\yco by-2
        \ifnum\yco>-240
        \repeat
        \put(\xco,\yco){$\kbl$}}
        \xco=158
        \yco=12
        \put(\xco,\yco){$\ktr$}
        \advance\yco by-1
        {\loop
        \put(\xco,\yco){$\kcr$}
        \advance\yco by-2
        \ifnum\yco>-240
        \repeat
        \put(\xco,\yco){$\kbr$}}
        \put(-20,13){\tiny University of Maryland Elementary Particle
Physics University of Maryland Elementary Particle Physics University of
Maryland Elementary Particle Physics}
        \put(-20,-241.5){\tiny University of Maryland Elementary
Particle Physics University of Maryland Elementary Particle Physics
University of Maryland Elementary Particle Physics}
        \end{picture}
        \par\vskip-8mm}
\def\bordero{                                           
        \setlength{\unitlength}{1mm}
        \newcount\xco
        \newcount\yco
        \xco=-31
        \yco=12
        \begin{picture}(140,0)
        \put(\xco,\yco){$\ktl$}
        \advance\yco by-1
        {\loop
        \put(\xco,\yco){$\kclr}
        \advance\yco by-2
        \ifnum\yco>-240
        \repeat
        \put(\xco,\yco){$\kbl$}}
        \xco=151
        \yco=12
        \put(\xco,\yco){$\ktr$}
        \advance\yco by-1
        {\loop
        \put(\xco,\yco){$\kcr$}
        \advance\yco by-2
        \ifnum\yco>-240
        \repeat
        \put(\xco,\yco){$\kbr$}}
        \put(-20,12){\ooo bacdefghidfghghdhededbihdgdfdfhhdheidhdhebaaahjhhdahb
a

hgdedge
   hgfdiehhgdigicba}
        \put(-20,-241.5){\ooo ababaighefdbfghgeahgdfgafagihdidihiidhiagfedhadbf
d

ecdcdfa
   gdcbhaddhbgfchbgfdacfediacbabab}
        \end{picture}
        \par\vskip-8mm}
\def\headpic{                                           
        \indent
        \setlength{\unitlength}{.4mm}
        \thinlines
        \par
        \begin{picture}(29,16)
        \put(165,16){\line(1,0){4}}
        \put(170,16){\line(1,0){4}}
        \put(180,16){\line(1,0){4}}
        \put(175,0){\line(1,0){4}}
        \put(180,0){\line(1,0){4}}
        \put(185,0){\line(1,0){4}}
        \put(169,0){\line(0,1){16}}
        \put(170,0){\line(0,1){16}}
        \put(179,0){\line(0,1){16}}
        \put(180,0){\line(0,1){16}}
        \put(184,0){\line(0,1){16}}
        \put(185,0){\line(0,1){16}}
        \put(169,16){\oval(8,32)[bl]}
        \put(170,16){\oval(8,32)[br]}
        \put(179,0){\oval(8,32)[tl]}
        \put(185,0){\oval(8,32)[tr]}
        \end{picture}
        \par\vskip-6.5mm
        \thicklines}
\def\title#1#2#3#4{\border\headpic {\hbox to\hsize{#4 \hfill UMDEPP #3}}\par
        \begin{center} \vglue .5in {\large\bf #1}\\[.6in]
        {#2}\\[.1in] {\it Department of Physics and Astronomy}\\
        {\it University of Maryland, College Park, MD 20742}\\[1.5in]
        {\bf ABSTRACT}\\[.1in] \end{center} \begin{quotation}}  
\def\Title#1#2#3#4#5#6#7{\border\headpic
        {\hbox to\hsize{#7 \hfill UMDEPP #6}}\par
        \begin{center} \vglue .4in {\large\bf #1}\\[.4in]
        {#2}\\[.1in] {\it Department of Physics and Astronomy}\\
        {\it University of Maryland, College Park, MD 20742}\\[.1in]
        {#3}\\[.1in] {\it {#4}}\\ {\it {#5}}\\[.5in] {\bf ABSTRACT}\\[.1in]
        \end{center} \begin{quotation}}                 
\def\endtitle{\end{quotation}\newpage}                  

\def\sect#1{\bigskip\medskip \goodbreak \noindent{\bf {#1}} \nobreak \medskip}

  \def\pp{{\mathchoice
          {
              \kern 1pt%
              \raise 1pt
              \vbox{\hrule width5pt height0.4pt depth0pt
                    \kern -2pt
                    \hbox{\kern 2.3pt
                          \vrule width0.4pt height6pt depth0pt
                          }
                    \kern -2pt
                    \hrule width5pt height0.4pt depth0pt}%
                    \kern 1pt
           }
            {
              \kern 1pt%
              \raise 1pt
              \vbox{\hrule width4.3pt height0.4pt depth0pt
                    \kern -1.8pt
                    \hbox{\kern 1.95pt
                          \vrule width0.4pt height5.4pt depth0pt
                          }
                    \kern -1.8pt
                    \hrule width4.3pt height0.4pt depth0pt}%
                    \kern 1pt
            }
            {
              \kern 0.5pt%
              \raise 1pt
              \vbox{\hrule width4.0pt height0.3pt depth0pt
                    \kern -1.9pt  
                    \hbox{\kern 1.85pt
                          \vrule width0.3pt height5.7pt depth0pt
                          }
                    \kern -1.9pt
                    \hrule width4.0pt height0.3pt depth0pt}%
                    \kern 0.5pt
            }
            {
              \kern 0.5pt%
              \raise 1pt
              \vbox{\hrule width3.6pt height0.3pt depth0pt
                    \kern -1.5pt
                    \hbox{\kern 1.65pt
                          \vrule width0.3pt height4.5pt depth0pt
                          }
                    \kern -1.5pt
                    \hrule width3.6pt height0.3pt depth0pt}%
                    \kern 0.5pt
            }
        }}

  \def\mm{{\mathchoice
                       {
                             \kern 1pt
               \raise 1pt    \vbox{\hrule width5pt height0.4pt depth0pt
                                  \kern 2pt
                                  \hrule width5pt height0.4pt depth0pt}
                             \kern 1pt}
                       {
                            \kern 1pt
               \raise 1pt \vbox{\hrule width4.3pt height0.4pt depth0pt
                                  \kern 1.8pt
                                  \hrule width4.3pt height0.4pt depth0pt}
                             \kern 1pt}
                       {
                            \kern 0.5pt
               \raise 1pt
                            \vbox{\hrule width4.0pt height0.3pt depth0pt
                                  \kern 1.9pt
                                  \hrule width4.0pt height0.3pt depth0pt}
                            \kern 1pt}
                       {
                           \kern 0.5pt
             \raise 1pt  \vbox{\hrule width3.6pt height0.3pt depth0pt
                                  \kern 1.5pt
                                  \hrule width3.6pt height0.3pt depth0pt}
                           \kern 0.5pt}
                       }}

\begin{document}


\border\headpic {\hbox to\hsize{January 1996 \hfill UMDEPP 96-64}}\par
\begin{center}
\vglue .4in
{\large\bf  Tuning the RADIO to the \\
Off-Shell 2D Fayet Hypermultiplet Problem
\footnote{Research supported
by NSF grant \# PHY-93-41926 and by NATO Grant CRG-93-0789} 
  }\\[.2in]
S. James Gates, Jr.\footnote{gates@umdhep.umd.edu} and 
Lubna Rana\footnote{lubna@umdhep.umd.edu} \\[.1in]
{\it Department of Physics\\
University of Maryland at College Park\\
College Park, MD 20742-4111, USA} 
\\[2.8in]

{\bf ABSTRACT}\\[.1in]
\end{center}
\begin{quotation}

We show via use of the RADIO technique that an off-shell (4,0) version of 
the hypermultiplet, in the form first proposed by Fayet, exists and contains 
28 - 28 component fields. The off-shell structure uncovered is found to 
include a chiral truncation of the ``generalized 2D, N = 4 tensor multiplet 
formalism'' proposed by Ketov. The (4,0) theory is extended to an
off-shell 56 - 56 component field (4,4) theory with the addition of a 
minimal (4,0) minus spinor multiplet together with (4,0) auxiliary 
multiplets. We propose that our final result gives a solution to a
twenty year-old 2D supersymmetry problem in the physics literature.

\endtitle

\sect{I. Introduction}

Twenty years ago, P. Fayet \cite{A} posed an intellectual ``Gordian Knot'' 
when he presented the on-shell hypermultiplet.  This was the first example
of a component level formulation which possesses 4D, N = 2 supersymmetry.
The ``knot'' to be unraveled is the problem of finding a complete set 
of auxiliary fields to accompany the on-shell spectrum that he found.
The supersymmetry algebra in his original formulation closes on some of 
the fields {\underline {only}} with the use of equations of motion.  
The physical plus auxiliary fields would then allow an off-shell realization 
of the 4D, N = 2 supersymmetry algebra without the need to use equations of 
motion to close the algebra.  A better known but similar problem is that of 
finding auxiliary fields for 4D, N = 4 supersymmetric Yang-Mills systems.  
Both of these are examples of the general ``off-shell N-extended supersymmetry 
problem.''  

The off-shell problem for 4D, N = 1 supersymmetry was essentially solved 
by the introduction of Salam-Strathdee superspace and superfields \cite{B}. 
For higher values of N, the problem remains unsolved because of the need 
to impose kinematic constraints\footnote{In a sense, kinematic constraints 
may be called Bianchi identities.} on the superfields in order to obtain 
irreducible or minimal off-shell representations. In fact, for a given 
off-shell representation, it is not even clear how many superfields are 
needed!  Up until now, the choice of these kinematic constraints has been 
dependent on guesswork, luck, etc. as opposed to being a 
science.  Within the last year however, we have seen hints from some 
developments within the study of 1D supersymmetric representation theory 
\cite{H} that these required differential equations are actually determined 
by an algebraic structure we denote by ${\cal G} {\cal R}( d, \, {\rm N})$. 
In particular, we have suggested that all supersymmetry representations 
(both on-shell and off-shell) are isomorphic to representations of this 
algebra. We have also found that there exist certain duality-like 
transformations that act on the representations of the algebra.

The off-shell supersymmetry problem even extends all the way
to superstring and heterotic string theory \cite{B1}! In this last
regard, it is conceivable that an increase in our understanding of the
auxiliary fields may even help provide some insight into the fundamental 
problem facing covariant superstring field theory, namely ``What is the
stringy extension of the equivalence principle of general relativity?''

The situation with off-shell supersymmetry is highly unsatisfactory. One of 
the main reasons is that the auxiliary fields are closely related to the 
different types of dynamics to which the propagating fields are subject.
This was recently illustrated with the example of the  4D, N = 1 WZNW 
term \cite{C}. As well, this situation is frustrating.  This has led to 
responses that can be placed into three categories; (a.) unreasonable 
proposals to go off-shell \cite{D}, (b.) theorems which ``prove'' off-shell 
representations suitable for actions are not possible with a finite number 
of auxiliary fields \cite{E} and (c.) off-shell representations suitable 
for actions with an infinite number of auxiliary fields \cite{F}.

The last one of these permits the construction of off-shell actions that 
propagate the same physical degrees of freedom as the on-shell theory.
Along these lines, we have long suspected that it should be possible to
truncate these theories so that only a finite number of auxiliary fields
will suffice to describe an off-shell theory.  This was based on a
little known observation of 4D, N = 1 scalar multiplet theory \cite{G}
where it was shown that it is possible to formulate the 4D, N = 1 
scalar multiplet so that it requires an infinite number of auxiliary 
fields to describe an off-shell action.  Although this is possible, it is 
terribly inconvenient to use such a formulation. After all, it is far simpler 
to use one of the off-shell representations that possess a finite number
of auxiliary fields.

Until recently, the almost total lack of understanding of off-shell 
N-extended supersymmetry representations even extended down to the 
simplest ones, the spinning particles.  Some time ago, we resolved to
undertake a comprehensive study of the off-shell problem within the
context of 1D theories \cite{H}. The most interesting outcome 
of this study has been the new point of view it has provided on the 
off-shell supersymmetry problem.  In fact, we have found both new on-shell 
representations \cite{I} and off-shell representations \cite{J} due to 
the use of new tools unearthed in our study. Perhaps the most unexpected 
of these new tools is the RADIO technique \cite{I}. The RADIO technique 
permits the derivation of new on-shell and off-shell representations by 
starting from a D-dimensional, N-extended theory reducing (R) to 1D, 
performing certain ``automorphic duality'' (AD) transformations, integrating 
additional 1D representations (I) and then oxidizing (O) back up to 
a different D-dimensional, N-extended theory.

Although, Fayet's original puzzle has not been solved before, there does
exists an off-shell 4D, N = 2 representation that should be closely
related to the putative one for the hypermultiplet.  Some time ago,
it was shown that a ``relaxed hypermultiplet'' \cite{K} exists. This
is an excellent candidate upon which to apply the RADIO. This will be
the purpose of this presentation. By the end of this present work we 
will propose that the 2D version of this puzzle has a surprisingly simple 
solution.  There exists a 28 - 28 off-shell (4,0) hypermultiplet that 
possesses exactly the on-shell limit required by the Fayet's original 
hypermultiplet formulation. We will also find that there exists a
28 - 28 off-shell (4,0) minus spinor multiplet\footnote{There is a
widespread misnomer regarding minus spinor multiplets. In the literature
these are \newline ${~~~~~}$ often called (0,4) multiplets.  This is 
technically incorrect because the multiplets are still only  
\newline ${~~~~~}$ representations of (4,0) supersymmetry.} that has 
the correct structure so that when it is added to the (4,0) hypermultiplet, 
the two together describe an off-shell full (4,4) hypermultiplet
with 56 - 56 components.

\sect{II. (4,4) Relaxed Hypermultiplet $\to$ RADIO $\to$ (4,0)
Original \\ ${~~~~}$ Hypermultiplet}

The work of Howe, Stelle and Townsend \cite{K} can be recalled by introducing
three real superfields $(S , \, L_{i j} , \,  L_{i j k l} )$ 
$$
S ~=~ {\Bar S} ~~~~,~~~~  L_{i j} ~=~ C_{i k} C_{j l} {\Bar L}^{k l} ~~~~,
~~~~ L_{i j k l}  ~=~ C_{i r} C_{j s} C_{k t} C_{l u} {\Bar L}^{r s t u}
~~~~, \eqno(2.1) $$
that satisfy two types of differential equations.  These are summarized in 
the table below,
\begin{center}
\renewcommand\arraystretch{1.2}
\begin{tabular}{|c|c| }\hline
${\rm Kinematic~Constraints}$
& ${\rm~Equations~of~Motion}$  \\ \hline
\hline
$ D_{\a ( i | } L_{| j k l m) } ~=~ 0 $ &  $ ~~~D_{\a i} S ~=~ i \frac 23
C^{j k} D_{\a j} L_{i k } ~~~ $ \\ \hline
$D_{\a ( i | } L_{| j k ) } ~=~ C^{l m} D_{\a l} L_{i j k m} 
$ &  $ L_{i j k l } ~=~ 0$ \\ \hline
$C^{i j} D_{\a i} D_{\b j} S ~=~ 0  $ &   ${~~~~~~~}$ \\ \hline
$  [\, D_{\a i} ~,~ {\Bar D}_{\dot \b}{}^i \,] S ~=~ 0$ &  ${~~~~~}$  \\ \hline
\end{tabular}
\end{center}
\vskip.1in
\centerline{{\bf Table I}}
The kinematic constraints above are an example of the guesswork
mentioned in the introduction.  The equations of motion are derived from 
an action of the form
$$
 {\cal S}_{{\rm RHM}} ~=~  \int d^4 x \, d^4 \z \, d^4 {\bar \z} ~
~ [ ~ (\lambda_{\a}{}^i \rho^{\a} {}_i ~+~
{\Bar \lambda}_{\dot \a}{}_i {\Bar \rho}^{\dot \a} {}^i ) ~+~ L^{ijkl} X_{ijkl}
~] ~~~,
\eqno(2.2) $$
where variation with respect to the pre-potential ${\rho}^{\a} {}_i$ leads 
to the first equation of motion and variation with respect to $X_{ijkl}$ 
leads to the second equation of motion.  Finally, a complete enumeration 
of the component fields is given by
$$
( L_{i j}, \, L_{i j k l} , \, \l_{\a i} , \, \psi_{\a i j k} , \, M_{i j}
, \, N , \, G_a , \, V_{a \, i} {}^j , \, \xi_{\a i} ~) ~~~~,
\eqno(2.3) $$
for those contained in ${\rho}^{\a} {}_i$ and as well
$$
( S , \, \psi_{\a i} , \, K_{i j} , \, A_{a \, i} {}^j , \, \xi_{\a
i j k} , \, C_{ijkl} ~) ~~~~,
\eqno(2.4) $$
for those in $X_{ijkl}$. The engineering dimensions of these fields are
given by
\begin{center}
\renewcommand\arraystretch{1.2}
\begin{tabular}{|c|c| }\hline
${\rm Component~Fields}$
& ${\rm~2D~Engineering~Dimension}$  \\ \hline
\hline
$ L_{i j}, \, L_{i j k l} , \,S $ &  $ ~~~0~~~ $ \\ \hline
$ \l_{\a i}, \,  \, \psi_{\a i} , \, \psi_{\a i j k}  $ &  $ ~~~1/2~~~ $ \\ \hline
$ K_{i j} , \, M_{i j} , \, N , \, G_a , \, V_{a \, i} {}^j , \, A_{a \, i} {}^j 
$ &  $ ~~~1~~~ $ \\ \hline
$ \xi_{\a i}, \,   \xi_{\a i j k}  $ &  $ ~~~3/2~~~ $ \\ \hline
$ C_{i j k l}  $ &  $ ~~~2~~~ $ \\ \hline
\end{tabular}
\end{center}
\centerline{{\bf Table II}}
The first step of the RADIO is to reduce this to 1D.  This will increase the
number of supersymmetries to four.  If we instead reduce to a 2D 
heterotic (4,0) formulation this is equivalent to the 1D theory. Under this 
reduction each of the pre-potentials splits into two distinct representations. We will
call the first of these, obtained after the reduction, the 2D (4,0) 
RHM (relaxed hypermultiplet) theory. The component fields of the 
2D (4,0) RHM representation are given by,
$$
( L_{i j}, \, L_{i j k l} , \, \l_{+ i} , \, \psi_{+ i j k} , 
\, G_{\pp} , \, \, G_{\mm} , \, V_{\pp \, i} {}^j , \, V_{\mm \, i} {}^j ,
\, \xi_{- i} ~) ~~~~,
\eqno(2.5) $$
and as well
$$
( S , \, \psi_{+ i} , \, A_{\pp \, i} {}^j , \,  A_{\mm \, i} {}^j , \,
\xi_{- i j k} , \, C_{ijkl} ~) ~~~~.
\eqno(2.6) $$

We call the second representation a non-minimal minus spinor multiplet
(NM${}^2$SM).  The component fields of the NM${}^2$SM representation are
$$
( \l_{- i} , \, \psi_{- i j k} , \, M_{i j} , \, N , \, s , \, p , 
\, s_{i} {}^j , \,  \, p_{i} {}^j , \, \xi_{+ i} ~) ~~~~,
\eqno(2.7) $$
and as well
$$
( \psi_{- i} ,  \, {\Tilde s}_i {}^j , \, {\Tilde p}_i {}^j , \,
K_{i j} , \, \xi_{+ i j k}  ~)  ~~~~.
\eqno(2.8) $$
If we start with $G_a$ in four dimensions, under reduction to 2D we
have $G_a ~\to ~ (s, \, p, \, G_{\pp} , $ $\, G_{\mm} \,)$.  The resulting
fields split between the two 2D (4,0) multiplets.  The two fields
$(s, \, p \,)$ go to the NM${}^2$SM theory.  The two fields $(G_{\pp} 
, \, G_{\mm} \,)$ go to the 2D (4,0) RHM theory. Similar observations
apply to $A_{a \, i} {}^j$ and $V_{a \, i} {}^j$. One other point of
convenience is to note that the components $s$ and $p$ can be
combined into two complex fields $u = s \, + \, i p$ and ${\Bar u} = s 
\, - \, i p$.

We next apply a Klein flip operator to the NM${}^2$SM theory.  The net
effect of this operator is that it multiplies every field in (2.7) by
a lower $+$-index and each field in (2.8) by a lower $-$-index.  The
Klein flip is one of a number of duality-type transformations that
have been found to exist for 1D N-extended supersymmetric theories \cite{H}.
It also has the interesting effect that it turns scalar multiplets into
spinor multiplets and vice-versa.  The multiplet we obtain after this 
AD map has a spectrum that consists of two sub-multiplets.  The component 
fields of the first, with two field strengths ${\cal A}_i$ 
and ${\cal A}_{i j k}$, has the structure
$$
( A_i , \, A_{i j k} , \, {\Bar \p}^- , \, {\r}^- , \, {\Bar \l}^- 
{}_{i j} , \, {\r}^- {}_i {}^j , \, P_{\pp \, i} ~) ~~~~,
\eqno(2.9) $$
and the second has the field strength ${\cal B}_{\mm \, i}$
along with the component field content,
$$
( P_{\mm \, i} , \, {\Bar \chi}^+ {}_{i j} , \, \b^+ {}_i {}^j , \, 
B_{i j k} ~) ~~~~.
\eqno(2.10) $$
All fields with more than one SU(2) index are totally symmetrical when
those indices are at the same height. This is a consequence of applying
the Klein flip operator to the RHM theory.

\sect{III.  An Off-Shell Version of the (4,0) Original Hypermultiplet}

The basic superfields that contain all of the component fields of the 
off-shell (4,0) version of the original hypermultiplet are ${\cal A}_i$, 
${\cal A}_{i j k}$ and ${\cal B}_{\mm i}$.  The spectrum of the fields
in (2.9) and (2.10) imply that the superfields satisfy the kinematic 
constraints,
$$ 
D_{+ ( i} {\cal A}_{j )} ~=~ C^{k l} D_{+ k} {\cal A}_{i j l} ~~~~,~~~~
D_{+ (i } {\cal A}_{j k l )} ~=~ 0  ~~~~, \eqno(3.1) 
$$
$$ {\Bar D}^j_+ {\cal A}_i ~-~ \frac 12 \d_i 
{}^j {\Bar D}^k_+ {\cal A}_k ~ = ~ C^{j k} {\Bar D}^l_+ {\cal A}_{i k l}
~~~~,~~~~ C_{m (i|} {\Bar D}^m_+ {\cal A}_{| j k l )} ~=~ 0 ~~~~.
\eqno(3.2) $$
$$ 
C^{i j} D_{+  i} {\cal B}_{\mm j} ~=~ 0 ~~~~,~~~~ {\Bar D}^i_+
{\cal B}_{\mm i} ~=~ 0 ~~~~. \eqno(3.3) 
$$
These kinematic constraints may be compared with those of the (4,0) RHM 
obtainable from Table I.  Examination of the first two lines above reveals 
that these equations are the (4,0) chiral projection of one of the special 
cases of the ``generalized 2D, N = 4 tensor multiplet formalism'' 
proposed by Ketov \cite{L}.  The component fields are obtained 
via the following projections
$$
 A_i  ~\equiv~ {\cal A}_i | ~~~~,~~~~ A_{i j k} ~\equiv~ {\cal A}_{i j k} |
~~~~,~~~~ {\Bar \p}^- ~\equiv~ - \frac 12 C^{i j} D_{+ i} {\cal A}_j | ~~~~,
~~~~   \r^- ~ \equiv ~ \frac 12 {\Bar D}^i_+ {\cal A}_i | ~~~~, 
$$
$$ 
{\Bar \l}^- {}_{i j} ~\equiv ~ \frac 12 D_{+ ( i} {\cal A}_{j )} | ~~~~,
~~~~ \r^- {}_i {}^j ~ \equiv ~ [\, {\Bar D}^j_+ {\cal A}_i ~-~ \frac 12 \d_i 
{}^j {\Bar D}^k_+ {\cal A}_k \, ] | ~~~~, ~~~~ P_{\pp \, i} ~\equiv ~ \frac 12 
{\Bar D}^j_+ D_{+ ( i} {\cal A}_{j )} | ~~~~,  
$$
$$ 
P_{\mm \, i} ~\equiv ~ [ \, {\cal B}_{\mm \, i} ~+~ i \pa_{\mm} {\cal A}_i 
\,] | ~~~,~~~ {\Bar \chi}^+ {}_{i j} ~\equiv ~ \frac 12 [ \, D_{+ ( i |} ( \,
{\cal B}_{\mm | j ) } ~+~ i \frac 12 \pa_{\mm} {\cal A}_{ | j ) } \, ) \, ] |
~~~,$$ 
$$\b^+ {}_i {}^j ~ \equiv ~ [\, {\Bar D}^j_+ ( \, {\cal B}_{\mm i } ~+~ i 
\frac 12 \pa_{\mm} {\cal A}_{ i } \, ) ~-~ i \frac 14 \d_i {}^j {\Bar D}^k_+ 
{\cal A}_k  \, ]  | ~~~~, $$
$$
B_{i j k} ~\equiv ~ [ \, C_{l \, ( i |} {\Bar D}^l_+ D_{+ | j |} {\cal B}_{\mm 
| k )} ~-~ \pa_{\pp} \pa_{\mm} {\cal A}_{i j k} \, ] | ~~~~.  \eqno(3.4)
$$
We do not explicitly write the supersymmetry variations of these fields
since these are easily calculated from $\d_Q \, = \, \e^{+ i} D_{+ i}
\, + \, {\Bar \e}^+ {}_i {\Bar D}_+ {}^i $. (See the appendix.)

There are two separate superinvariants that we can form utilizing the
fields in (3.4). This should be expected since similar structures
existed for the RHM which was our starting point. The first 
supersymmetrically invariant Lagrangian is of the form,
$$ \eqalign{
{\cal L}_1 ~=~ &(\pa_{\mm} A_i \,) ( \pa_{\pp} {\Bar A}^i \, ) ~+~
i \frac 12 {\Bar \p}^- \pa_{\mm} \p^- ~+~ i \frac 12 {\Bar \r}^- 
\pa_{\mm} \r^- \cr
& -~ \frac 16 (\pa_{\mm} A_{i j k} \,) ( \pa_{\pp} {\Bar A}^{i j k} \, ) 
~+~ i \frac 32 {\Bar P}_{\pp} {}^i \, (\pa_{\mm} A_i \,) ~-~
i \frac 32 {P}_{\pp} {}_i \, (\pa_{\mm} {\Bar A}^i \,) \cr
& -~ i \frac 34 {\Bar \l}^- {}_{i j} \pa_{\mm} {\l}^- {}^{i j} ~
 -~ i \frac 34 {\Bar \r}^- {}_i {}^j \pa_{\mm} {\r}^- {}_j {}^i ~~~~.
} \eqno(3.5) $$
A second supersymmetrically invariant Lagrangian takes the form
$$ \eqalign{ {~~~~~}
{\cal L}_2 ~=~ & \frac 32 (~ {\Bar P}_{\pp} {}^i \, {P}_{\mm} {}_i
 ~+~ {\Bar P}_{\mm} {}^i \, {P}_{\pp} {}_i ~) ~-~ \frac1{12} (~ 
{\Bar B}^{i j k} A_{i j k} ~+~ {\Bar A}^{i j k} B_{i j k} ~) \cr
& \,-~ \frac 34 (~ {\Bar \r}^- {}_i {}^j {\b}^+ {}_j {}^i ~-~
{\r}^- {}_i {}^j {\Bar \b}^+ {}_j {}^i ~ ) ~-~ \frac 34
(~ {\l}^- {}^{i j} {\Bar \chi}^+ {}_{i j} ~-~ {\Bar \l}^- {}_{i j} 
{\chi}^+ {}^{i j} ~ ) \cr
& \,+~ \frac 16 (\pa_{\mm} A_{i j k} \,) ( \pa_{\pp} {\Bar A}^{i j k} \, ) 
~-~ i \frac 32 {\Bar P}_{\pp} {}^i \, (\pa_{\mm} A_i \,) ~+~
i \frac 32 {P}_{\pp} {}_i \, (\pa_{\mm} {\Bar A}^i \,) \cr
& \,+~ i \frac 34 {\Bar \l}^- {}_{i j} \pa_{\mm} {\l}^- {}^{i j} ~
 +~ i \frac 34 {\Bar \r}^- {}_i {}^j \pa_{\mm} {\r}^- {}_j {}^i 
~~~~.
} \eqno(3.6) $$
Using the Lagangian ${\cal L}^{(4,0)}_{\rm {OHM}} \equiv {\cal L}_1 + {\cal 
L}_2$, we find the final form of the action
$$ \eqalign{ {~~}
{\cal S}^{(4,0)}_{\rm {OHM}} ~=~  \int d^2 \s & \Big[ ~ (\pa_{\mm} A_i \,)
( \pa_{\pp} {\Bar A}^i \, ) ~+~ i \frac 12 {\Bar \p}^- \pa_{\mm} \p^- 
~+~ i \frac 12 {\Bar \r}^- \pa_{\mm} \r^- {~~~~} \cr
& +~ \frac 32 (~ {\Bar P}_{\pp} {}^i \, {P}_{\mm} {}_i ~+~ {\Bar P}_{\mm} 
{}^i \, {P}_{\pp} {}_i ~) ~-~ \frac1{12} (~ {\Bar B}^{i j k} A_{i j k} 
~+~ {\Bar A}^{i j k} B_{i j k} ~)  \cr
& -~  \frac 34 (~ {\Bar \r}^- {}_i {}^j {\b}^+ {}_j {}^i ~-~
{\r}^- {}_i {}^j {\Bar \b}^+ {}_j {}^i ~ ) ~-~  \frac 34
(~ {\l}^- {}^{i j} {\Bar \chi}^+ {}_{i j} ~-~ {\Bar \l}^- {}_{i j} 
{\chi}^+ {}^{i j} ~ ) ~ \Big] ~~~~.
} \eqno(3.7) $$
The component fields uniquely fix the unconstrained pre-potentials that
describe the two multiplets. The first pre-potential is given
by a dimension minus-one superfield $\Psi_{\mm \, i}$ (containing
the fields in (2.9)) and the second pre-potential is a dimension zero 
superfield $\Psi_{\mm \mm \, i j k}$ (containing the fields in 
(2.10)).

\sect{IV. (4,4) Relaxed Hypermultiplet $\to$ RADIO $\to$ 
(4,0) MSM-III Theory}

Thus far, we have solved half of the long standing problem of finding 
an off-shell formulation whose on-shell limit contains the hypermultiplet 
as first described by Fayet.  Our 2D (4,0) OHM theory satisfies this 
constraint. The solution to the complete problem of an off-shell 2D, N 
= 4 hypermultiplet requires the discovery of an additional 2D (4,0) 
minus spinor multiplet. The OHM theory in (2.9) and (2.10) is the analog 
of the RHM in (2.5) and (2.6). What is required is another version of the 
NM${}^2$SM multiplet. Surprisingly, application of the Klein flip operator 
to (2.5) and (2.6) does {\underline {not}} give the required version the 
NM${}^2$SM multiplet.

Another 1D duality-type transformation is the automorphic duality (AD) 
map.  Upon reducing the multiplets in (2.7) and (2.8) to one dimension, we 
can use the AD map upon the multiplets. When this is done appropriately,
we arrive at a different 2D (4,0) NM${}^2$SM multiplet. The spectrum of
the component fields is contained in three multiplets. The first
multiplet is composed of 4 - 4 fields
$$
(  {\Bar \p}^+ , \, {\r}^+ , \, F_i )
\eqno(4.1) $$ 
a second composed of 12 - 12 component fields
$$ ( {\Bar \l}^+ {}_{i j} , \, {\r}^+ {}_i {}^j ,
\, F_{i j k} , \,  \, G_i ~) ~~~~,
\eqno(4.2) $$
and a third multiplet also composed of the 12 - 12 component fields,
$$
( K_i , \, K_{i j k} , \, {\Bar \chi}^- {}_{i j} , \, \b^- {}_i {}^j 
 ~) ~~~~.
\eqno(4.3) $$
The first of these multiplets is actually one of the minimal minus
spinor multiplets (SM-III) described previously \cite{M}. The latter
multiplets are completely auxiliary and are a minus spinor multiplet
and scalar multiplet respectively.

These multiplets have a number of algebraically independent field strength
superfields; $ ( \, {\Bar {\P}} {}^+ , \, {\U}^+ \,) $, $(\, {\Bar {\P}} 
{}^+ {}_{i j} , \, {\U}^+ {}_i {}^j \, ) $ and $ ( \, {\cal H}_i , \, 
{\cal H}_{i j k} ) $. These field strengths satisfy a number of kinematic 
constraints,
$$
D_{+ i} {\Bar {\P}} {}^+ ~=~ {\Bar D}_+ {}^i \U^+ ~=~ D_{+ i} \U^+ ~-~ 
C_{i j} {\Bar D}_+ {}^j  {\Bar {\P}} {}^+ ~=~ 0 ~~~, $$
$$
D_{+ i} {\Bar {\P}} {}^+ {}_{j k}  ~=~ {\Bar D}_+ {}^i {\U}^+ {}_j {}^k ~=~ 
0 ~~~, ~~ C_{l \, j } D_{+ i} {\U}^+ {}_{k } {}^l ~=~ - \, C_{l \, i } 
{\Bar D}_+ {}^l {\Bar {\P}} {}^+ {}_{j k}  ~~~, $$
$$
C^{i j} D_{+ i} {\cal H}_j ~=~ {\Bar D}_+ {}^i {\cal H}_i ~=~0 ~~~,~~~$$
$$
D_{+ i} {\cal H}_{j k l} ~=~ C_{i ( j|}  D_{+ |k|} {\cal H}_{|l )}
~~~ , ~~~ {\Bar D}_+ {}^i {\cal H}_{j k l} ~=~ \d_{( j|} {}^i 
{\Bar D}_+ {}^m {\cal H}_{|k|} C_{|l) m} ~~~.
\eqno(4.4) $$

Similarly, the spectrum of component fields can be defined via projection
as
$$ 
{\Bar \p}^+ ~ \equiv ~ {\Bar \P}{}^+ | ~~~~, ~~~~ {\r}^+ ~ \equiv 
~ \U^+ | ~~~~, ~~~~ F_i  ~ \equiv ~ C_{i j} {\Bar D}_+ {}^j 
{\Bar \P}{}^+  |~~~~, 
$$
for the 4 - 4 multiplet and 
$$
{\r}^+ {}_i {}^j ~ \equiv ~ \U^+ {}_i {}^j | ~~~~,~~~~
~~~~ {\Bar \l}^+ {}_{i j} ~\equiv ~ {\Bar \P}{}^+ {}_{i j} |~~~~
, $$ 
$$F_{i j k}  ~ \equiv ~  \frac 14 C_{l \, ( i |} {\Bar D}_+ {}^l
{\Bar \P}{}^+ {}_{ |j k) } |~~~~, ~~~~ G_i ~ \equiv ~ \frac 13 [ ~ 
D_{+ j} \U^+ {}_i {}^j  ~ ] | ~~~~,
\eqno(4.5) $$
for one of the 12 - 12 component multiplets and
$$ 
K_i ~\equiv ~ {\cal H}_i | ~~~,~~~ K_{i j k} ~\equiv ~ {\cal 
H}_{i j k} |  ~~~, ~~~ {\Bar \chi}^- {}_{i j} ~ \equiv ~ \frac 12 
D_{+ ( i} {\cal H}_{ j ) } | ~~~,~~~ 
\b^- {}_i {}^j ~ \equiv ~  {\Bar D}_+ {}^j {\cal H}_i  | ~~~~,
\eqno(4.6) $$ 
for the remaining 12 - 12 component multiplet.

Once again there are two separate superinvariants that we can form 
utilizing the fields in (4.4) and (4.5). The first 
supersymmetrically invariant Lagrangian is of the form,
$$
{\Hat {\cal L}_1} ~=~   i \frac 12 {\Bar \p}^+ \pa_{\pp} \p^+ ~+~ i 
\frac 12 {\Bar \r}^+ \pa_{\pp} \r^+ ~+~ \frac 14 {\Bar F}^i F_i
~~~~.
 \eqno(4.7) $$
A second supersymmetrically invariant Lagrangians takes the form
$$ \eqalign{ {~~~~~}
{\Hat {\cal L}_2} ~=~ &  \frac 32 (~ {\Bar G}^i \, {K}_i ~+~ 
{\Bar K}^i \, G_i ~) ~-~ \frac 1{12}  (~ {\Bar K}^{i j k} F_{i j k} 
~+~ {\Bar F}^{i j k} K_{i j k} ~)  \cr
&{} -~ \frac 34 (~ {\l}^+ {}^{i j} {\Bar \chi}^- {}_{i j} ~-~ {\Bar 
\l}^+ {}_{i j} {\chi}^- {}^{i j} ~ ) ~-~ \frac 34   (~ {\Bar \r}^+ 
{}_i {}^j {\b}^- {}_j {}^i ~-~ {\r}^+ {}_i {}^j {\Bar \b}^- {}_j 
{}^i ~ ) ~~~~.
} \eqno(4.8) $$
We take the final Lagrangian to be a sum ${\cal L}^{(4,0)}_{{\rm {NM}}{}^2
{\rm {SM-III}}} \equiv {\Hat {\cal L}_1} + {\Hat {\cal L}_2}$ and this
yields a component action given by
$$ \eqalign{
{\cal S}^{(4,0)}_{\rm {NM}{}^2 {\rm {SM-III}}} ~=~ \int d^2 \s
 & \Big[ ~  i \frac 12 {\Bar \p}^+ \pa_{\pp} \p^+ ~+~ i \frac 12 {\Bar 
\r}^+ \pa_{\pp} \r^+ ~+~ \frac 14 {\Bar F}^i F_i   \cr
& +~ \frac 32 (~ {\Bar G}^i \, {K}_i ~+~ 
{\Bar K}^i \, G_i ~)  ~-~ \frac 1{12}  (~ {\Bar K}^{i j k} F_{i j k} 
~+~ {\Bar F}^{i j k} K_{i j k} ~)  \cr
&{} -~ \frac 34 (~ {\l}^+ {}^{i j} {\Bar \chi}^- {}_{i j} ~-~ {\Bar 
\l}^+ {}_{i j} {\chi}^- {}^{i j} ~ ) ~-~ \frac 34 (~ {\Bar \r}^+ 
{}_i {}^j {\b}^- {}_j {}^i ~-~ {\r}^+ {}_i {}^j {\Bar \b}^- {}_j 
{}^i ~ ) ~\Big] ~~~~.
} \eqno(4.9) $$
Once again the structure of the component fields fixes the pre-potential
that describe this theory. The fields in (4.1) are contained in
a dimension minus one pre-potential $\Psi_{\mm \mm i}$. The fields
of (4.2) are contained in two dimension minus one pre-potentials 
${\Hat \Psi}_{\mm \mm i}$ and ${\Hat \Psi}_{\mm \mm \, i j k}$.
Finally the fields of (4.3) are contained in two chiral dimension
one-half pre-potentials ${\Psi}_{- \, i j }$ and ${\Hat \Psi}_{- \, i}
{}^j$.

\sect{V. 2D (4,4) Off-shell OHM Theory}

It should be obvious that what we have derived in the last two sections
are the two different chiral parts of a single 2D (4,4) theory. Our
method of derivation, since it is actually one-dimensional, treats the
two chiral parts separately. The results of these derivations must
be integrated together into a single theory. We begin by noting that
an action of the form
$$
{\cal S}^{(4,4)}_{\rm {OHM}} ~=~ {\cal S}^{(4,0)}_{\rm {OHM}} ~+~
{\cal S}^{(4,0)}_{\rm {NM}{}^2 {\rm {SM-III}}} ~~~~, 
\eqno(5.1) $$
contains all the degrees of freedom to describe the full (4,4) theory.
Furthermore, if we interchange {\it {all}} $+$-signs with $-$-signs and
vice-versa, we obtain the same action.  So the sum in (5.1) has, in addition 
to (4,0) supersymmetry, a symmetry under parity reflections. This is
exactly the necessary and sufficient condition of a $(p,0)$ theory to 
imply the existence of a full $(p,p)$ supersymmetry. 

The derivation of the form of the (0,4) supersymmetry variations can
be done as follows. We first note that the complete spectrum of
component fields is given in (2.9), (2.10), (4,1), (4.2) and (4.3).
We next assert the existence of (0,4) operators $D_{- i}$ and ${\Bar 
D}_- {}^i$ acting on all component fields such that $D_{- i} {\cal 
S}^{(4,4)}_{\rm {OHM}} = 0 $.  The realization of $D_{- i}$ on all 
fields is determined by taking the realization of  $D_{+ i}$ on all 
fields and replacing {\underline {all}} plus signs by minus signs 
and vice-versa. 

Now having described how the (0,4) supersymmetry operators are derived,
in principle we can derive the realization of the full (4,4) supersymmetry 
algebra. Without giving explicit results, we note the form it
takes is given by
$$ [\, D_{+ \, i} ~,~ {\Bar D}_{+}^{~ j} \, \} ~=~ i 2 \d_i {}^j
\pa_{\pp} ~~,~~ [\, D_{- \, i} ~,~ {\Bar D}_{-}^{~ j} \, \} ~=~ i 2 
\d_i {}^j \pa_{\mm} ~~, ~~[\, D_{\pm \, i} ~,~  D_{\pm \, j} \, \} 
~=~ 0 ~~~, $$
$$ [\, D_{+ \, i} ~,~  D_{- \, j} \, \} ~=~  i 2\, {\cal Z}^{(1)} {}_{i j} 
~~~,~~~ [\, D_{+ \, i} ~,~ {\Bar D}_{-}^{~ j} \, \} ~=~ i 2\, {\cal Z}^{(
2)}{}_i {}^j  ~~~. \eqno(5.2) $$
In other words, the full (4,4) algebra requires the presence of
two complex central charges ${\cal Z}^{(1)}{}_{i j}$ and ${\cal 
Z}^{(2)}{}_i {}^j$
.

\sect{VI. Summary}

In this work we have used the RADIO technique to derive new results
for off-shell realizations of 2D (4,0) supersymmetry.  This powerfully
demonstrates the surprising capabilities inherent in the 1D formulation
of higher dimensional theories.  The fact, that we were able to start 
from the relaxed hypermultiplet and derive results for the Fayet
hypermultiplet suggests that there are many such connections to even
more non-minimal theories. In fact, our past experience \cite{M} with 
the minimal off-shell formulations taught us to expect the (4,0) theories
to come in several distinct versions.  The multiplet in (2.3) 
and (2.4) is a non-minimal extension of the SM-II theory (in the 
classification of (4,0) scalar multiplets of \cite{M}). The scalar multiplet 
in (2.9) and (2.10) is a non-minimal extension of SM-III. Similarly, the
spinor multiplet of (2.7) and (2.8) is a non-minimal extension of the 
MSM-I and/or MSM-II theories.

The occurrence of these non-minimal representations also has possible 
implications for the ADHM non-linear $\s$-model construction \cite{M1}.  
In our first investigation of the manifest realization of (4,0) 
supersymmetry in this class of models, we showed that it was not possible 
to construct such a theory using only {\underline {minimal}} and 
manifest (4,0) supersymmetry representations.  This led some \cite{N} 
to the premature conclusion that it is not possible to construct off-shell 
(4,0) ADHM models utilizing a finite number of auxiliary fields.  This 
conclusion was premature since no use of non-minimal representations was 
investigated.

We have long suspected that the apparently infinite number of auxiliary
fields that seemed to be required by harmonic superspace formulations
is an illusion. Our present work supports this view.  If this view is 
always true, then the harmonic superspace approach is simply an
expensive luxury (i.e. $112 << \infty$).  It will be important to 
derive our present results from harmonic superspace since this may 
permit the latter to be used as a method of derivation of off-shell 
formulations with a finite numbers of auxiliary fields in other theories 
(i.e. N = 4 supersymmetric Yang-Mills theory, etc.).  The harmonics 
introduced in the harmonic superspace approach are likely 
to ultimately be analogous to the extra ``Vainberg coordinate'' \cite{O} 
introduced by Novikov and Witten \cite{P} for the description of the WZNW 
term of QCD. The Vainberg construction is a convenient, not essential,
method for describing the WZNW action. Similarly, harmonic superspace 
may be a convenient way to derive off-shell formulations. We propose 
that there should exist projection operators for harmonic superspace 
that can be used to recover distinct off-shell theories with a finite 
number of auxiliary fields.

{\bf {Appendix A: $D$-Algebra Realization and Alternate Formulations}}

In this appendix, we give the complete and explicit realization of the 
$D_+$-operators on all fields. The realization of the $D_-$-operators is 
uniformly obtained by interchanging {\underline {all}} plus and minus 
signs below.
$$
D_{+ i} A_j ~=~ - C_{i j} {\Bar \p}^- ~+~ {\Bar \l}^- {}_{i j} ~~~,~~~
{\Bar D}_+ {}^i A_j ~=~ \d_j {}^i \r^- ~+~ \r^- {}_j {}^i ~~~,$$
$$
D_{+ i} {\Bar \l}^- {}_{j k} ~=~ 0  ~~~,~~~ {\Bar D}_+ {}^i \r^-
{}_j {}^k ~=~ 0  ~~~,~~~
D_{+ i} {\Bar \p}^- ~=~ 0  ~~~,~~~ {\Bar D}_+ {}^i \r^- ~=~ 0  ~~~, $$
$$  D_{+ i} \r^- ~=~ - [ \, 3 P_{\pp \, i} ~+~ i 2 \pa_{\pp} A_i \, ] ~~~,~~~
{\Bar D}_+ {}^i {\Bar \p}^- ~=~ C^{i j} [ \, 3 P_{\pp \, j} ~+~ i 2 
\pa_{\pp} A_j \, ] ~~~,  $$
$$ D_{+ i}  \r^- {}_j {}^k ~=~ 2 [ \, \d_i {}^k  P_{\pp \, j} ~-~
\frac 12 \d_j {}^k  P_{\pp \, i} \, ] ~+~ i \frac 23 C^{k l} \pa_{\pp} 
A_{i j l} ~~~,$$
$$ {\Bar D}_+ {}^i {\Bar \l}^- {}_{j k} ~=~ \d_{  ( j |} {}^i
 P_{\pp \, | k ) } ~-~ i \frac 23 C^{i l} \pa_{\pp} 
A_{j k l} ~~~,$$
$$  D_{+ i}  P_{\pp \, j} ~=~ - i \frac 23 \pa_{\pp} {\Bar \l}^- 
{}_{i j} ~~~,~~~ {\Bar D}_+ {}^i  P_{\pp \, j} ~=~ - i \frac 23 \pa_{\pp}
\r^- {}_j {}^i ~~~, $$
$$  D_{+ i} A_{j k l} ~=~ \frac 12 C_{i ( j |} {\Bar \l}^- {}_{| k l )}
~~~,~~~  {\Bar D}_+ {}^i  A_{j k l} ~=~ \frac 12 \d_{( j |} {}^i
\r^- {}_{| k |} {}^m C_{| l) m} ~~~,  \eqno(A.1) $$

$$ D_{+ i}  P_{\mm \, j} ~=~ {\Bar \chi}^+ {}_{i j} ~+~ i \frac 12
\pa_{\mm} {\Bar \l}^- {}_{i j} ~-~ i C_{i j} \pa_{\mm}  {\Bar \p}^- ~~~,
~~~  D_{+ i} {\Bar \chi}^+ {}_{j k} ~=~ 0  ~~~$$
$$  {\Bar D}_+ {}^i  P_{\mm \, j} ~=~ \b^+ {}_j  {}^i ~+~ i \frac 12
\pa_{\mm} \r^- {}_j {}^i ~+~ i \d_j {}^i \pa_{\mm} \r^- ~~~
,~~~ {\Bar D}_+ {}^i \b^+
{}_j {}^k ~=~ 0  ~~~, $$
$$ \eqalign{
{\Bar D}_+ {}^i {\Bar \chi}^+ {}_{j k} ~=~ & -~ \frac 23 \d_{( j |} 
{}^i \pa_{\pp} \pa_{\mm} A_{| k )} ~+~ 
i \frac 12 \d_{(j |} {}^i [ \, \pa_{\mm}
P_{\pp \, | k )} ~-~ \frac 43 \pa_{\pp} P_{\mm \, | k )} \, ] ~-~
\frac 13 C^{i l}  B_{j k l}   ~~~, }$$
$$  \eqalign{
D_{+ i} \b^+ {}_j {}^k ~= &-~ \frac 43  \pa_{\pp} \pa_{\mm} [ \, \d_i 
{}^k A_j ~-~ \frac 12 \d_j {}^k A_i \, ] ~+ ~ i \d_i {}^k [ \, \pa_{\mm} 
P_{\pp \, j} ~-~ \frac 43 \pa_{\pp} P_{\mm \, j} \, ] \cr 
&- ~ i \frac 12 \d_j {}^k [ \, \pa_{\mm} P_{\pp \, i} ~-~ \frac 43 
\pa_{\pp} P_{\mm \, i} \, ] ~+~ \frac 13 C^{k l}  B_{i j l}  ~~~, }$$
$$  D_{+ i} B_{j k l} ~=~ i C_{i ( j |} \pa_{\pp}
{\Bar \chi}^+ {}_{| k l )}  ~~~, ~~~ {\Bar D}_+ {}^i  B_{j k l} 
~=~ i \d_{( j |} {}^i \pa_{\pp} \b^+ {}_{| k |} {}^m  C_{| l) m} 
~~~,  \eqno(A.2) $$

$$  D_{+ i} {\Bar \p}^+ ~=~ 0 ~~~,~~~  {\Bar D}_+ {}^i \r^+ ~=~ 0 ~~~,
~~~ D_{+ i} \r^+ ~=~ F_i ~~~,~~~ {\Bar D}_+ {}^i {\Bar \p}^+ ~=~ - C^{i 
j} F_j ~~~, $$
$$  D_{+ i} F_j ~=~ i 2 C_{i j} \pa_{\pp} {\Bar \p}^+ ~~~,~~~
{\Bar D}_+ {}^i F_j ~=~  - i 2 \d_j {}^i \pa_{\pp} \r^+ ~~~,
 \eqno(A.3) $$

$$  D_{+ i} \r^+ {}_j {}^k ~=~ 2 (\, \d_i {}^k G_j ~-~ \frac 12
\d_j {}^k G_i \, ) ~+~ \frac 23 C^{k l} F_{i j l} ~~~,~~~ 
{\Bar D}_+ {}^i  \r^+ {}_j {}^k ~=~ 0 ~~~, $$
$$  D_{+ i} {\Bar \l}^+ {}_{j k} ~=~ 0 ~~~,~~~  {\Bar D}_+ {}^i 
{\Bar \l}^+ {}_{j k} ~=~ \d_{( j} {}^i G_{k )} ~-~ \frac 23 C^{i l}
F_{j k l} ~~~,$$
$$   D_{+ i}  G_j ~=~ - i \frac 23  \pa_{\pp} {\Bar \l}^+ {}_{i j}
~~~,~~~ {\Bar D}_+ {}^i  G_j ~=~ - i \frac 23  \pa_{\pp} \r^+ {}_j {}^i
~~~, $$
$$  D_{+ i} F_{j k l} ~=~ i \frac 12 C_{i (j |} \pa_{\pp} {\Bar \l}^+ 
{}_{| k l) }  ~~~, ~~~ {\Bar D}_+ {}^i F_{j k l} ~=~ i \frac 12 \d_{(j |}
{}^i \pa_{\pp} { \r}^+ {}_{| k |} {}^m C_{| l) \, m} ~~~. 
\eqno(A.4)     $$
 
$$D_{+ i}  K_j ~=~ {\Bar \chi}^- {}_{i j} ~~~,~~~ {\Bar D}_+ {}^i  
K_j ~=~ \b^- {}_j {}^i ~~~, ~~~
D_{+ i} {\Bar \chi}^- {}_{j k} ~=~ 0 ~~~, ~~~ {\Bar D}_+ {}^i  
\b^- {}_j {}^k ~=~ 0 ~~~, $$
$$
D_{+ i} \b^- {}_j {}^k ~=~ - i \frac 43 (\, \d_i {}^k  \pa_{\pp} K_j ~-~ 
\frac 12 \d_j {}^k  \pa_{\pp} K_i \, ) ~+~ i \frac 13 C^{k l} 
\pa_{\pp}K_{i j l} ~~~, $$
$$ {\Bar D}_+ {}^i {\Bar \chi}^- {}_{j k} ~=~- i \frac 23 \d_{( j} {}^i 
\pa_{\pp} K_{k )} ~-~ i\frac 13 C^{i l} \pa_{\pp} K_{j k l} ~~~,$$
$$  D_{+ i} K_{j k l} ~=~ C_{i (j |} {\Bar \chi}{}^- {}_{| k l) }  ~~~, 
~~~ {\Bar D}_+ {}^i K_{j k l} ~=~ \d_{(j |} {}^i { \b}^- {}_{| k |} 
{}^m C_{| l) \, m} ~~~. 
\eqno(A.5)     $$

Finally we note that the formulation described above lends itself to 
other re-formulations of the off-shell 2D (4,4) OHM
theory. Under the following field re-definitions
$P_{\pp \, i} ~\to ~ i \pa_{\pp} P_i ,~ P_{\mm \, i} ~\to~ i {\bo} 
{}^{-1}  \pa_{\mm} B_i $ a different local theory exists.  It has three 
bosonic field strengths ${\cal A}_i$,  ${\cal P}_i$ and ${\cal P}_{i j k}$
containing the component fields 
$$
( A_i , \, {\Bar \p}{}^- , \, {\r}^- ~) ~~~, ~~~ ( A_{i j k} ,  \, P_i ,\, 
{\Bar \l}^- {}_{i j} , \, {\r}^- {}_i {}^j  ~) ~~~~,
\eqno(A.6) $$
and as well spinorial field strength tensors ${\cal C}^+ {}_{i j}$ and 
${\cal B}^+ {}_i {}^j$ containing the component field content,
$$
({\Bar \chi}^+ {}_{i j} , \, \b^+ {}_i {}^j , \, 
B_{i j k} , \,  B_i   ~) ~~~~.
\eqno(A.7) $$
The component form of the action is obtained by applying the field
re-definition described above to (3.7). This version of the (4,0) OHM 
has a structure like that of (4.1), (4.2) and (4.3) where the first 
multiplet in (A.6) replaces the minus spinor multiplet. The spectrum 
of fields in (4.1) - (4.3) together with those of (A.6) and (A.7) has 
the interesting feature that they can all be interpreted as arising from 
2D, N = 2 chiral superfields: $\Phi_i$, ${\Hat \Phi}_i$, ${\Tilde \Phi
}_i$, ${\Hat \Phi}_{i j k}$ and  ${\Tilde \Phi}_{i j k}$.

One other alternative formulation utilizes the re-definitions $B_{i j k} 
~\to ~ - i \pa_{\pp} P_{\mm \, i j k}$, $A_{i j k} ~\to~ - i {\bo} 
{}^{-1}  \pa_{\mm} P_{\pp \, i j k} $. Under this re-definition, the 
multiplets of (2.9) and (2.10) are replaced by
$$
( A_i , \, {\Bar \p}^- , \, {\r}^- , \, {\Bar \l}^- {}_{i j} , \, 
{\r}^- {}_i {}^j , \, P_{\pp \, i} , \, P_{\pp \, i j k} ~) ~~,~~
( P_{\mm \, i} , \, P_{\mm \, i j k} , \, {\Bar \chi}^+ {}_{i j}
, \, \b^+ {}_i {}^j  ~) ~~~~.
\eqno(A.8) $$

\newpage


\begin{thebibliography}{66}

\bibitem{A} P. Fayet, Nucl. Phys. {\bf {B113}} (1976) 135.

\bibitem{B} A. Salam and J. Strathdee, {\it {Intro. to Supersymmetry}},
Fortschr. Phys. {\bf {26}} (1978) 960.

\bibitem{H} S. J. Gates, Jr. and L. Rana, Univ. of Maryland Preprint, 
UMDEPP 93-074, in progress.

\bibitem{B1} M. B. Green and J. Schwarz, Nucl. Phys. {\bf {B198}} (1982),
252, ibid. 441; Phys. Lett. {\bf {136B}} (1984) 367.
 
\bibitem{C} S. J. Gates, Jr., Univ. of Maryland Preprint, UMDEPP 
96-19, to appear in Phys. Lett. 

\bibitem{D} M. Sohnius, K. Stelle and P. West, Phys. Lett. {\bf {92B}} 
(1980) 123;  B.E.W. Nilsson, Class. Quant. Grav. {\bf {3}} (1986) L41.
 
\bibitem{E} M. Ro\v cek and W. Siegel, Phys. Lett. {\bf {105B}} (1981) 275;
V. Rivelles and J. Taylor, J. Phys. A. Math. Gen. {\bf {15}} (1982) 163.


\bibitem{F} A. Galperin, E. Ivanov, S. Kalitzin, V. Ogievetsky and
E. Sokatchev, Class. Quant. Grav. {\bf {1}} (1984) 469.

\bibitem{G} S. J. Gates, Jr. and W. Siegel, Nucl. Phys. {\bf {B187}} (1981) 
389, S. J. Gates, Jr. and B. B. Deo, Nucl. Phys. {\bf {B254}} (1984) 187.

\bibitem{I} S. J. Gates, Jr. and L. Rana, Phys. Lett. {\bf {342B}} 
(1995) 132.

\bibitem{J} S. J. Gates, Jr. and L. Rana, Univ. of Maryland Preprint, 
UMDEPP 96-38, to appear in Phys. Lett.

\bibitem{K} P. S. Howe, K. S. Stelle and P. K. Townsend, Nucl. 
Phys. {\bf {B214}} (1983) 519, {\it idem.} {\bf {B236}} (1984) 125; 
H. D. Dahmen and Marculescu, Nucl. Phys. {\bf {B292}} (1987) 298.

\bibitem{L} S. V. Ketov, Int.~J.~Mod.~Phys.~{\bf {A3}} (1988) 703.

\bibitem{M} S. J. Gates, Jr. and L. Rana, Phys. Lett. {\bf {345B}} 
(1994) 233.

\bibitem{M1} E. Witten, J. Geom. Phys. {\bf {15}} (1995) 215. 

\bibitem{N} A. Galperin and E. Sokatchev, Nucl.Phys. {\bf {B452}} (1995)
431 and Johns Hopkins Univ. preprint, JHU-TIPAC-94021, hep-th/9504124; 
N. Lambert, Univ. of Camb. preprint DAMTP R/95/43, hep-th/9508039.

\bibitem{O} M. M. Vainberg, {\it {Variational Methods for the Study of
Nonlinear Operators}} (Holden Day, San Francisco, 1964), 135.

\bibitem{P} S. P. Novikov,  Soviet Math. Dokl. {\bf {24}} (1981) 222; 
idem. Uspekhi Mat. Nauk. {\bf {37}} (1982) 3 (in Russian); E. Witten, 
Nucl. Phys. {\bf {B223}} (1983) 422; idem. Commun. Math. Phys. {\bf 
{92}} (1984) 455.

\end{thebibliography}
\end{document}